\documentclass[11pt,titlepage]{article} %
\usepackage{times}
\usepackage{latexsym}
\usepackage{amsfonts}
\usepackage{amssymb}
\mathcode`\0="0030      %
\mathcode`\1="0031
\mathcode`\2="0032
\mathcode`\3="0033
\mathcode`\4="0034
\mathcode`\5="0035
\mathcode`\6="0036
\mathcode`\7="0037
\mathcode`\8="0038
\mathcode`\9="0039

\makeatletter
\def\@begintheorem#1#2{\trivlist\item[\hskip\labelsep{\bf #1\ #2}]}
\def\foobarpt{\textfont\z@\tenrm 
  \scriptfont\z@\ninrm \scriptscriptfont\z@\sevrm
\textfont\@ne\tenmi \scriptfont\@ne\ninmi \scriptscriptfont\@ne\sevmi
\textfont\tw@\tensy \scriptfont\tw@\ninsy \scriptscriptfont\tw@\sevsy
\textfont\thr@@\tenex \scriptfont\thr@@\tenex \scriptscriptfont\thr@@\tenex
\def\unboldmath{\everymath{}\everydisplay{}\@nomath\unboldmath
          \textfont\@ne\tenmi 
          \textfont\tw@\tensy \textfont\lyfam\tenly
          \@boldfalse}\@boldfalse
\def\boldmath{\@ifundefined{tenmib}{\global\font\tenmib\@mbi\@magscale1\global
        \font\tensyb\@mbsy \@magscale1\global\font
         \tenlyb\@lasyb\@magscale1\relax\@addfontinfo\@xiipt
              {\def\boldmath{\everymath
                {\mit}\everydisplay{\mit}\@prtct\@nomathbold
                \textfont\@ne\tenmib \textfont\tw@\tensyb 
                \textfont\lyfam\tenlyb\@prtct\@boldtrue}}}{}\@xiipt\boldmath}%
\def\prm{\fam\z@\tenrm}%
\def\pit{\fam\itfam\tenit}\textfont\itfam\tenit \scriptfont\itfam\ninit
   \scriptscriptfont\itfam\sevit
\def\psl{\fam\slfam\tensl}\textfont\slfam\tensl 
     \scriptfont\slfam\tensl \scriptscriptfont\slfam\tensl
\def\pbf{\fam\bffam\tenbf}\textfont\bffam\tenbf 
   \scriptfont\bffam\ninbf \scriptscriptfont\bffam\ninbf 
\def\ptt{\fam\ttfam\tentt}\textfont\ttfam\tentt
   \scriptfont\ttfam\nintt \scriptscriptfont\ttfam\nintt 
\def\psf{\fam\sffam\tensf}\textfont\sffam\tensf
    \scriptfont\sffam\tensf \scriptscriptfont\sffam\tensf
\def\psc{\@getfont\psc\scfam\@xiipt{\@mcsc\@magscale1}}%
\def\ly{\fam\lyfam\tenly}\textfont\lyfam\tenly 
   \scriptfont\lyfam\ninly \scriptscriptfont\lyfam\sevly
 \@setstrut \rm}

\newcommand{\singlespacing}{\let\CS=
\@currsize\renewcommand{\baselinestretch}{1}\tiny\CS}
\newcommand{\singlespacingplus}{\let\CS=
\@currsize\renewcommand{\baselinestretch}{1.25}\tiny\CS}
\newcommand{\doublespacing}{\let\CS=
\@currsize\renewcommand{\baselinestretch}{1.75}\tiny\CS}
\newcommand{\draftspacing}{\let\CS=
\@currsize\renewcommand{\baselinestretch}{2.0}\tiny\CS}

\newcommand{\niceonespacing}{\let\CS=\@currsize\renewcommand{\baselinestretch}{1.1}\tiny\CS}
\newcommand{\nicetwospacing}{\let\CS=\@currsize\renewcommand{\baselinestretch}{1.2}\tiny\CS}
\newcommand{\nicethreespacing}{\let\CS=\@currsize\renewcommand{\baselinestretch}{1.3}\tiny\CS}
\newcommand{\singlespacingplusplus}{\let\CS=\@currsize\renewcommand{\baselinestretch}{1.35}\tiny\CS}
\newcommand{\nicefivespacing}{\let\CS=\@currsize\renewcommand{\baselinestretch}{1.5}\tiny\CS}
\newcommand{\nicesixpacing}{\let\CS=\@currsize\renewcommand{\baselinestretch}{1.6}\tiny\CS}
\newcommand{\nicefoospacing}{\let\CS=\@currsize\renewcommand{\baselinestretch}{1.15}\tiny\CS}

\makeatother

\makeatletter
\def\@cite#1#2{[#1\if@tempswa , #2\fi]}
\makeatother

\newcommand\seq{\subseteq}
\renewcommand\.{\cdot}
\newcommand\<{\langle}
\renewcommand\>{\rangle}

\newcommand\Lora{\ \Longrightarrow \ }

\newcommand{\sigmastar}{\mbox{$\Sigma^\ast$}}
\newcommand\equalsdef{\stackrel{\mbox{\protect\scriptsize df}}{=}}
\newcommand\tweak{\hspace*{1pt}}

\newcommand\xor{\mbox{$\sym$}}

\def\sym{\Delta}

\newcommand\N{{\rm I\!N}}

\newcommand\p{\mbox{\rm P}}
\newcommand\fp{\mbox{\rm FP}}
\newcommand\np{\mbox{\rm NP}}

\newcommand\conp{\mbox{\rm coNP}}

\newcommand{\psel}{\mbox{\rm P-Sel}}
\newcommand{\ppoly}{\mbox{\rm P/poly}}

\newtheorem{theorem}{Theorem}%
\newtheorem{corollary}[theorem]{Corollary}
\newtheorem{lemma}[theorem]{Lemma}

\newtheorem{fact}[theorem]{Fact}

\newtheorem{definition}[theorem]{Definition}

\newtheorem{thm}[theorem]{Theorem}
\newtheorem{lem}[theorem]{Lemma}

\newtheorem{cor}[theorem]{Corollary}

\begin{document}

\title{
Boolean Operations, Joins, and the Extended Low Hierarchy
} 

\author{
\ \ \ \ {\em  Lane A. Hemaspaandra\/} \protect\thanks{
Supported in part
by grants NSF-INT-9513368/DAAD-315-PRO-fo-ab, 
NSF-CCR-8957604, 
NSF-INT-9116781/JSPS-ENG-207, and 
NSF-CCR-9322513. Work done in part while visiting 
Friedrich-Schiller-Universit\"at Jena and the 
Tokyo Institute of Technology.}\protect\\
\ \ \ \ Dept.\ of Computer Science\protect\\
\ \ \ \ University of Rochester\protect\\
\ \ \ \ Rochester, NY 14627, USA\\\\
\and
\ \ \ \ {\em  Zhigen Jiang\/} \protect\thanks{
Supported in part by a postdoctoral fellowship from the Chinese
Academy of Sciences, and by grant NSF-CCR-8957604. 
Work done in part while visiting the 
University of Rochester and while at
McMaster University.  Current address: 
Suite 1100, 
123 Front Street,
Toronto, Ontario, M5J 2M3 Canada.}\protect\\
\ \ \ \ Institute of Software\protect\\
\ \ \ \ Chinese Academy of Sciences\protect\\
\ \ \ \ Beijing 100080, China\\\\
\and
{\em  J\"org Rothe\/} \protect\thanks{
Supported in part by a DAAD research visit grant, and by grants
NSF-INT-9513368/DAAD-315-PRO-fo-ab, 
NSF-CCR-9322513, and 
NSF-CCR-8957604.  
Work done in part while visiting the University of Rochester. 
Contact via email: {\tt rothe@informatik.uni-jena.de}.}\protect\\
Institut f\"ur Informatik\protect\\
Friedrich-Schiller-Universit\"at Jena\protect\\
07743 Jena, Germany\\
\and
{\em  Osamu Watanabe\/} \protect\thanks{
Supported in part by grant NSF-INT-9116781/JSPS-ENG-207.  
Work done in part while visiting the 
University of Rochester.}\protect\\
Dept.\ of Computer Science\protect\\
Tokyo Institute of Technology\protect\\
Tokyo 152, Japan\\
} %

\newcount\hour  \newcount\minutes  \hour=\time  \divide\hour by 60
\minutes=\hour  \multiply\minutes by -60  \advance\minutes by \time
\def\mmmddyyyy{\ifcase\month\or Jan\or Feb\or Mar\or Apr\or May\or Jun\or Jul\or
  Aug\or Sep\or Oct\or Nov\or Dec\fi \space\number\day, \number\year}
\def\hhmm{\ifnum\hour<10 0\fi\number\hour :%
  \ifnum\minutes<10 0\fi\number\minutes}
\def\Draft{{\it Draft of \mmmddyyyy}}

\date{}

\typeout{WARNING:  BADNESS used to supress reporting!  Beware!!}
\hbadness=3000%
\vbadness=10000 %

\setcounter{footnote}{0}
{\singlespacing
{\maketitle}
}

{\singlespacing

\begin{center}
{\large\bf Abstract}
\end{center}

\begin{quotation}
  
\noindent 
We prove that the join of two sets may actually fall into a lower
level of the extended low hierarchy~\cite{bal-boo-sch:j:low} than
either of the sets.  In particular, there exist sets that are not in
the second level of the extended low
hierarchy, $\mbox{EL}_2$, yet their join {\em
is\/} in $\mbox{EL}_2$.  That is, in terms of extended lowness, the
join operator can lower complexity. Since in a strong intuitive sense the
join does not lower complexity, our result suggests that the extended
low hierarchy is unnatural as a complexity measure.  We also study the
closure properties of $\mbox{EL}_2$ and prove that $\mbox{EL}_2$ is
not closed under certain Boolean operations. To this end, we establish
the first known (and optimal) $\mbox{EL}_2$ lower bounds for certain
notions generalizing Selman's
P-selectivity~\cite{sel:j:pselective-tally}, which may be regarded as
an interesting result in its own right.

\end{quotation}
}

\section{Introduction}

The low hierarchy~\cite{sch:j:low} provides a yardstick
to measure the complexity of sets that are known to be in NP but that
are seemingly neither in P nor NP-complete.  In order to extend this
classification beyond NP, the extended low
hierarchy~\cite{bal-boo-sch:j:low} has been introduced (see the
surveys~\cite{koe:c:survey-low,hem:j:yardstick}).  
An informal way of describing the intuitive nature of 
these hierarchies might be the 
following:
A set $A$ that is placed in the $k$th level of the low or the
extended low hierarchy contains no more information than the
empty set relative to the computation of a $\Sigma_k^p$ machine
(see~\cite{mey-sto:c:reg-exp-needs-exp-space,sto:j:poly} for the
definition of the $\Sigma$ levels of the polynomial hierarchy), either
because $A$
is so chaotically organized that a $\Sigma_k^p$ machine is not able to
extract useful information from~$A$, or because $A$ is so simple 
that it has no useful information to offer a 
$\Sigma_k^p$ machine.\footnote{%
\protect\singlespacing%
We stress that this is a very loose and informal description.
In particular, for the case of the extended low hierarchy, it 
would be more accurate to say:
A set $A$ that is placed in the $(k+1)$st level of the 
extended low hierarchy, $k>1$, is such that 
$\np^A$
contains no more information than 
${\rm SAT} 
\oplus L$ relative to the computation of a $\Sigma_{k}^p$ machine.}
The low and extended low hierarchies have
been very thoroughly investigated in many papers
(see,
e.g.,~\cite{sch:j:low,ko-sch:j:circuit-low,bal-boo-sch:j:low,sch:j:gi,sch:j:pr-low,ko:j:separating-low-high,all-hem:j:low,ami-bei-gas:j-subm:uni,koe:j:locating,she-lon:j:extended-low,hem-nai-ogi-sel:j:refinements}).
In light of the informal
intuition given above---that classifying the level in the extended low
hierarchy of a problem or a class gives insight into the amount of
polynomial-hierarchy computational power needed to make access to the problem 
or the class redundant---one main motivation for the study of the 
extended low hierarchy is 
to understand which natural complexity classes
and problems easily extend the power of the polynomial hierarchy and
which do not.  Among the important natural classes and problems that
have been carefully classified in these terms are 
the Graph Isomorphism Problem (which in fact is known to be low),
bounded probabilistic polynomial time (BPP), 
approximate polynomial time (APT), 
the class of complements of sets having Arthur-Merlin games (coAM), 
the class of sparse and co-sparse sets, 
the P-selective sets, and
the class of sets having polynomial-size circuits (P/poly).
Another motivation for the study of the low and
extended low hierarchies is 
to relate their
properties to other complexity-theoretic concepts.  For instance,
Sch\"oning showed that the existence of an NP-complete set (under
any ``reasonable'' reducibility) in the low hierarchy implies a
collapse of the polynomial hierarchy~\cite{sch:j:low}.
Among the most important recent results about extended lowness
are Sheu and Long's result that the extended low hierarchy is a strictly 
infinite hierarchy~\cite{she-lon:j:extended-low} and K\"obler's optimal
location of P/poly in the extended low hierarchy~\cite{koe:j:locating}.   
In this note, we seek to further
explore the structure of the extended low hierarchy by studying
its interactions with such operations as the join.
In particular, we prove properties of $\mbox{EL}_2$ with regard to its
interaction with the join and with Boolean operations.
Our results add to the body of 
evidence that extended lowness does not provide a natural, intuitive
measure of complexity.

In light of the many ways in which extended lowness captures certain
concepts of low information content (such as all sparse sets and certain
reduction closures of the sparse sets---e.g., the Turing closure of the class 
of sparse sets, which is known to be equal to P/poly) as well as 
certain concepts
of ``almost'' feasible computation (such as BPP, APT, and
P-selectivity, etc.), it might be tempting to assume that extended lowness
would provide a reasonable measure of complexity in the sense that a
problem's property of being extended low indicates that this problem 
is of ``low'' complexity. However, in
Section~\ref{sec:join}, we will
prove that {\em the join operator can lower
difficulty as measured in terms of extended lowness\/}:
There exist sets that are not in
$\mbox{EL}_2$, yet their join is in $\mbox{EL}_2$.  
Since in a strong intuitive sense the
join does not lower complexity, our result suggests that, if one's intuition
about complexity is---as is natural---based on reductions,
then the extended low hierarchy is not a natural measure
of complexity.  Rather, it is a measure that is related
to the difficulty of information extraction, and it is 
in flavor quite orthogonal to more traditional notions
of complexity. 
That is, our result sheds light on the orthogonality of
``complexity in terms of reductions'' versus ``difficulty in terms of 
non-extended-lowness.'' 
In fact, our result is possible only since
the second 
level of the extended low hierarchy is not closed under
polynomial-time many-one reductions (this non-closure
is known, see~\cite{all-hem:j:low}, and it
also follows as a corollary of our result). 

In Section~\ref{sec:boolean}, we apply the technique developed in the
preceding section to prove that the second level of the extended low
hierarchy is not closed under the Boolean operations intersection,
union, exclusive-or, and equivalence.
Our result will follow from the proof
of another result, which establishes the first known (and optimal)
EL$_2$ lower bounds for generalized selectivity-like classes (that
generalize Selman's class of P-selective
sets~\cite{sel:j:pselective-tally}, denoted P-Sel) such as the
polynomial-time membership-comparable sets introduced by
Ogihara~\cite{ogi:j:comparable} and the multi-selective sets
introduced by 
Hemaspaandra et 
al.~\cite{hem-jia-rot-wat:t:multiselectivity}. These results
sharply contrast with the known result that all P-selective sets are
in~EL$_2$ and they are thus interesting in their own right.

\section{Extended Lowness and the Join Operator}
\label{sec:join}

The low hierarchy and the extended low hierarchy are defined as follows.

\begin{definition}
\label{def:low}
\begin{enumerate}
\item {\cite{sch:j:low}}\quad 
  For each $k\geq 1$, define \mbox{$\mbox{Low}_k
  \equalsdef \{ L\in \np \mid \Sigma_k^{p,\,L} = \Sigma_k^p\}$}.

\item {\cite{bal-boo-sch:j:low}}\quad  For each $k\geq
  2$, define EL$_k \equalsdef \{ L \mid \Sigma_k^{p,\,L} =
  \Sigma_{k-1}^{p,\, L \oplus \mbox{\scriptsize SAT}}\}$, where SAT is
  the set of all satisfiable Boolean formulas.
\end{enumerate}
\end{definition}

For sets $A$ and $B$, their join, $A\oplus B$, is $\{0x \mid x\in
A\}\cup\{1x \mid x\in B\}$.  Theorem~\ref{thm:el2-join} below
establishes that the join operator can lower the difficulty measured 
in terms of extended lowness.  At first glance, this might seem paradoxical.
After all, every set that $\leq_{m}^{p}$-reduces\footnote{%
\protect\singlespacing%
For sets $X$ and $Y$, $X \leq_{m}^{p} Y$ 
if and only if there is a polynomial-time
computable function $f$ such that \mbox{$X = \{ x \mid f(x) \in Y\}$}.
}
to a set $A$ or $B$ also reduces to
$A\oplus B$, and thus 
intuition strongly suggests that $A\oplus B$ must be at
least as hard as $A$ and $B$, as most complexity lower bounds (e.g.,
NP-hardness) are defined in terms of reductions. However, extended
lowness merely measures the opacity
of a set's internal organization, and
thus Theorem~\ref{thm:el2-join} is not paradoxical. Rather,
Theorem~\ref{thm:el2-join} highlights the orthogonality of
``complexity in terms of reductions'' and ``difficulty in terms of
non-extended-lowness.'' Indeed, note Corollary~\ref{cor:el2-join},
which was first observed by Allender and 
Hemaspaandra (then Hemachandra)~\cite{all-hem:j:low}.
We interpret Theorem~\ref{thm:el2-join} as evidence that extended lowness
is not an appropriate, natural complexity measure 
with regard to even very simple
operations such as the join. 

\begin{thm}
\label{thm:el2-join} \quad
There exist sets $A$ and $B$ such that $A \not\in \mbox{\rm EL}_{2}$
and $B \not\in \mbox{\rm EL}_{2}$, and yet \mbox{$A\oplus B \in
  \mbox{\rm EL}_{2}$}.
\end{thm}

Lemma~\ref{lem:el2-join} below will be used in the upcoming proof of
Theorem~\ref{thm:el2-join}. First, we fix some notations.  Fix the
alphabet $\Sigma = \{ 0,1 \}$.  Let $\Sigma^*$ denote the set of all
strings over~$\Sigma$.  For each set $L \seq \Sigma^*$, $L^{=n}$
($L^{\leq n}$) is the set of all strings in $L$ having length $n$
(less than or equal to $n$), and $\|L\|$ denotes the cardinality of
$L$.  Let $\Sigma^{n}$ be a shorthand for $(\Sigma^*)^{=n}$.  Let
$\leq_{\mbox{\protect\scriptsize lex}}$ denote the standard
quasi-lexicographical ordering on~$\sigmastar$.  The census function
of a set $L$ is defined by $\mbox{\em census\/}_{L}(0^n) = \| L^{\leq
  n} \|$.  $L$ is said to be sparse if there is a polynomial $p$ such
that for every~$n$, $\mbox{\em census\/}_{L}(0^n) \leq p(n)$.  Let SPARSE
denote the class of all sparse sets. For each class $\tweak{\cal C}$ of
sets over ${\rm \Sigma}$, define $\mbox{co}\tweak{\cal C} \equalsdef
\{L\,|\,\overline{L}\in {\cal C}\}$. Let $\N$ denote the set of
non-negative integers. To encode a pair of integers, we use a one-one,
onto, polynomial-time computable pairing function, $\<\.,\.\> : \N
\times \N \rightarrow \N$, that has polynomial-time computable
inverses.  FP denotes the class of polynomial-time computable
functions.  We shall use the shorthand NPTM to refer to
``nondeterministic polynomial-time Turing machine.''  For an NPTM $M$
(an NPTM $M$ and a set $A$, respectively), $L(M)$ ($L(M^A)$) denotes
the set of strings accepted by $M$ (relative to $A$).

\begin{lem}
\label{lem:el2-join} \quad
If $F$ is a sparse set and 
$\mbox{\em census\/}_{F} \in \fp^{F\oplus \mbox{\scriptsize SAT}}$,
then $F \in \mbox{\rm EL}_{2}$.
\end{lem}

\noindent 
{\bf Proof.} \quad 
Let $L \in \np^{\mbox{\scriptsize NP}^{F}}$ via NPTMs $N_1$ and $N_2$,
i.e., $L = L(N_{1}^{L(N_{2}^{F})})$. Let $q(n)$ be a polynomial
bounding the length of all queries that can be asked in the run of
$N_{1}^{L(N_{2}^{F})}$ on inputs of length $n$. Below we describe an
NPTM $N$ with oracle $F\oplus \mbox{SAT}$:

On input $x$, $|x| = n$, $N$ first computes $\mbox{\em census\/}_{F}(0^i)$
for each relevant length $i \leq q(n)$, and then guesses all sparse
sets up to length $q(n)$. Knowing the exact census of $F$, $N$ can use
the $F$ part of its oracle to verify whether the guess for $F^{\leq
  q(n)}$ is correct, and rejects on all incorrect paths. On the
correct path, $N$ uses itself, the SAT part of its oracle, and the
correctly guessed set $F^{\leq q(n)}$ to simulate the computation of
$N_{1}^{L(N_{2}^{F})}$ on input $x$.

Clearly, $L(N^{F\oplus \mbox{\scriptsize SAT}}) = L$\@. Thus,
$\np^{\mbox{\scriptsize NP}^{F}} \seq \np^{F\oplus \mbox{\scriptsize
    SAT}}$, i.e., $F \in \mbox{EL}_{2}$.~\hfill$\Box$

\medskip

\noindent 
{\bf Proof of Theorem~\ref{thm:el2-join}.} \quad 
$A  \equalsdef  \bigcup_{i\geq 0} A_i$ and 
$B  \equalsdef  \bigcup_{i\geq 0} B_i$ are
constructed in stages. In order to show $A \not\in \mbox{EL}_{2}$ and
$B \not\in \mbox{EL}_{2}$ it suffices to ensure in the construction
that $\np^A \not\seq \conp^{A \oplus \mbox{\scriptsize SAT}}$ and
$\np^B \not\seq \conp^{B \oplus \mbox{\protect\scriptsize SAT}}$ (and thus,
\mbox{$\np^{\mbox{\protect\scriptsize NP}^{A}} \not\seq \np^{A \oplus
  \mbox{\scriptsize SAT}}$} and \mbox{$\np^{\mbox{\protect\scriptsize NP}^{B}}
\not\seq \np^{B \oplus \mbox{\protect\scriptsize SAT}}$}).

\smallskip

Define function $t$ 
inductively by $t(0)  \equalsdef  2$ and
$t(i)  \equalsdef  2^{2^{2^{t(i-1)}}}$ for $i \geq 1$. Let 
$\{ N_i \}_{i\geq 1}$ be a fixed enumeration of all coNP
oracle machines having the property that 
the runtime of each $N_i$ is independent of the oracle and 
each machine appears infinitely 
often in the enumeration. Define 
\[
L_A  \equalsdef  \{ 0^{t(i)}\mid (\exists j \geq 1 )\, [ i = \< 0,j \>
\,\wedge\, \| A \cap \Sigma^{t(i)} \| \geq 1 ] \},
\]
\[
L_B  \equalsdef  \{ 0^{t(i)}\mid (\exists j \geq 1 )\, [ i = \< 1,j \>
\,\wedge\, \| B \cap \Sigma^{t(i)} \| \geq 1 ] \}.
\]
Clearly, $L_A \in \np^A$ and $L_B \in \np^B$\@.  In stage $i$ of the
construction, at most one string of length $t(i)$ will be added to $A$
and at most one string of length $t(i)$ will be added to $B$ in order
\begin{description}
\item[(1)] to ensure $L(N_j^{A_i \oplus \mbox{\scriptsize SAT}}) \neq
  L_A$ if $i = \<0,j\>$ (or $L(N_j^{B_i \oplus \mbox{\scriptsize
      SAT}}) \neq L_B$, respectively, if \mbox{$i = \<1,j\>$}), and 

\item[(2)] to encode an easy to find string into $A$ if $i = \<1,j\>$ (or
  into $B$ if $i = \<0,j\>$) indicating whether or not some string has
  been added to $B$ (or to $A$) in (1).
\end{description}  
Let $A_{i-1}$ and $B_{i-1}$ be the content of $A$ and
$B$ prior to stage $i$. Initially, let $A_0 = B_0 = \emptyset$. Stage
$i$ is as follows: 

\smallskip

First assume $i = \< 0,j \>$ for some $j\geq 1$.  If it is the case
that no path of $N_j^{A_{i-1} \oplus \mbox{\scriptsize
    SAT}}(0^{t(i)})$ can query all strings in $\Sigma^{t(i)} -
\{0^{t(i)}\}$ and $N_j^{A_{i-1} \oplus \mbox{\scriptsize
    SAT}}(0^{t(i)})$ cannot query any string of length $t(i+1)$
(otherwise, just skip this stage---we will argue later that the
diagonalization still works properly), then simulate $N_j^{A_{i-1}
  \oplus \mbox{\scriptsize SAT}}$ on input $0^{t(i)}$.  If it rejects
(in the sense of coNP, i.e., if it has one or more rejecting
computation paths), then fix some rejecting path and let $w_i$ be the
smallest string in $\Sigma^{t(i)} - \{0^{t(i)}\}$ that is not queried
along this path, and set \mbox{$A_i := A_{i-1} \cup \{ w_i \}$} and 
\mbox{$B_i := B_{i-1} \cup \{ 0^{t(i)} \}$}. 
Otherwise (i.e., if $0^{t(i)} \in
L(N_j^{A_{i-1} \oplus \mbox{\scriptsize SAT}})$), set \mbox{$A_i :=
A_{i-1}$} and \mbox{$B_i := B_{i-1}$}. The case of $i = \< 1,j \>$ is
analogous: just exchange $A$ and~$B$.  This completes the construction
of stage~$i$.

\smallskip

Since each machine $N_i$ appears infinitely often in our
enumeration and as the $t(i)$ are strictly increasing, it is clear
that for only a finite number of the $N_{i_1}, N_{i_2}, \ldots $
that are the same machine as $N_i$ can it happen that stage $i_k$ must
be skipped (in order to ensure that $w_{i_k}$, if needed to
diagonalize against $N_{i_k}$, indeed exists, or that the construction
stages do not interfere with each other), and thus each machine $N_i$
is diagonalized against eventually.  This proves that $A \not\in
\mbox{EL}_{2}$ and $B \not\in \mbox{EL}_{2}$. Now observe that
$A\oplus B$ is sparse and that $\mbox{\em census\/}_{A\oplus B} \in
\fp^{A\oplus B}$.
Indeed, 
$$\mbox{\em census\/}_{A\oplus B}(0^n) = 
2( \| A \cap \{ 0, 00, \ldots , 0^{n-1} \} \| +
   \| B \cap \{ 0, 00, \ldots , 0^{n-1} \} \|).$$
Thus, by Lemma~\ref{lem:el2-join}, $A\oplus B \in \mbox{EL}_{2}$.~\hfill$\Box$

\begin{cor}
\label{cor:el2-join}
{\rm \cite{all-hem:j:low}} \quad
$\mbox{\rm EL}_{2}$ is not closed under $\leq_{m}^{p}$-reductions.
\end{cor}

In contrast to the extended low hierarchy, every level of the
low hierarchy within NP is clearly closed under
$\leq_{m}^{p}$-reductions.  Thus, the low hierarchy analog of
Theorem~\ref{thm:el2-join} cannot hold.

\begin{fact} \quad
$(\forall k \geq 0)\, (\forall A,B)\, [(A \not\in \mbox{\rm Low}_{k} \,\vee\,
B \not\in \mbox{\rm Low}_{k}) \Lora A\oplus B \not\in \mbox{\rm Low}_{k} ]$.
\end{fact}

\noindent
{\bf Proof.} \quad
Assume $A\oplus B \in \mbox{Low}_{k}$. Since for all sets
$A$ and $B$, $A \leq_{m}^{p} A\oplus B$ and $B \leq_{m}^{p} A\oplus B$,
the closure of $\mbox{Low}_{k}$ under $\leq_{m}^{p}$-reductions implies
that both $A$ and $B$ are in $\mbox{Low}_{k}$.~\hfill$\Box$

\medskip

One of the most interesting open questions related to the results
presented in this note is whether the join operator also can {\em
  raise\/} the difficulty measured in terms of extended lowness. 
That is, do
there exist sets $A$ and $B$ such that \mbox{$A \in \mbox{EL}_{k}$}
and \mbox{$B \in \mbox{EL}_{k}$}, and yet \mbox{$A\oplus B \not\in
  \mbox{EL}_{k}$} for, e.g., $k=2$? Or is the second level of
the extended low hierarchy (and more generally, are {\em all\/}
levels of this hierarchy) closed under join? Regarding potential
generalizations of our result, we conjecture that 
Theorem~\ref{thm:el2-join} can be generalized to higher levels of the
extended low hierarchy. Such a result, to be sure, would probably
require some new technique such as a clever modification of the
lower-bound technique for constant-depth Boolean circuits developed by
Yao, H{\aa}stad, and Ko
(see, e.g., \cite{has:j:circuits,ko:j:separating-low-high}).

\section{EL{\boldmath$_2$} is not Closed Under Certain Boolean Connectives}
\label{sec:boolean}

In this section, we will prove that the second level of the extended
low hierarchy is {\em not\/} closed under the Boolean connectives
union, intersection, exclusive-or, or equivalence.  
We will do so by combining the
technique of the previous section with standard techniques of
constructing P-selective sets. To this end, we first seek to improve
the known EL$_2$ lower bounds of P/poly, the well-studied class of
sets having polynomial-size circuits~\cite{kar-lip:c:nonuniform}. To
wit, we will show that certain generalizations of the class of
P-selective sets, though still contained in
P/poly~\cite{ogi:j:comparable,hem-jia-rot-wat:t:multiselectivity}, are
not contained in~EL$_2$.  As interesting as this result may be in its
own right, its proof will even provide us with the means required to
show the above-mentioned main result of this section: EL$_2$ is not
closed under certain Boolean connectives (and indeed P-selective
sets can be used to witness the non-closure). This extends the main
result of Hemaspaandra
and Jiang~\cite{hem-jia:j:psel}, namely 
that P-Sel is not closed
under those Boolean connectives.

Let us first recall the following generalizations of Selman's
P-selectivity. Ogihara introduced the P-membership comparable
sets~\cite{ogi:j:comparable} and the present paper's authors
(\cite{hem-jia-rot-wat:t:multiselectivity}, see also~\cite{rot:phd})
introduced the notion of multi-selectivity as defined in
Definition~\ref{def:multisel}.

\begin{definition}
\label{def:p-mc}
\cite{ogi:j:comparable}
\quad
Fix a positive integer~$k$.
A function $f$ is called a $k$-membership comparing function
for a set~$A$ if and only if for every $w_1,\ldots ,w_m$
with $m \geq k$,
\begin{eqnarray*}
f(w_1,\ldots ,w_m)\in \{0,1\}^m & \mbox{and} & 
(\chi_A(w_1),\ldots ,\chi_A(w_m)) \neq f(w_1,\ldots ,w_m),
\end{eqnarray*}
where $\chi_A$ denotes the characteristic function of~$A$.  If in
addition $f \in \fp$, $A$ is said to be polynomial-time $k$-membership
comparable. Let P-mc$(k)$ denote the class of all polynomial-time
$k$-membership comparable sets.
\end{definition}

We can equivalently (i.e., without changing the class) require in the
definition that \mbox{$f(w_1,\ldots ,w_m) \neq (\chi_A(w_1),\ldots
  ,\chi_A(w_m))$} must hold only if the inputs \mbox{$w_1,\ldots
  ,w_m$} happen to be {\em distinct}.  This is true because if there
are $r$ and $t$ with $r \neq t$ and $w_r = w_t$, then $f$ simply
outputs a length $m$ string having a ``0'' at position $r$ and a ``1''
at position~$t$.

\begin{definition} 
\label{def:multisel}
Fix a positive integer~$k$.
  Given a set~$A$, a function $f \in \fp$ is said to be an
  $\mbox{\rm S}(k)$-selector for~$A$ if and only if $f$ satisfies the
  following property: For each set of 
  distinct input strings $y_1,\ldots ,y_n$,
\begin{enumerate}
\item $f(y_1,\ldots ,y_n) \in \{ y_1,\ldots ,y_n\}$, and

\item if $\| A \cap \{ y_1,\ldots ,y_n\} \| \geq k$, then
  $f(y_1,\ldots ,y_n) \in A$.
\end{enumerate}
The class of sets having an $\mbox{\rm S}(k)$-selector is denoted by
$\mbox{\rm S}(k)$. 
\end{definition}

It is easy to see that $\mbox{P-mc}(1) = \p$ and \mbox{$\mbox{S}(1) =
  \psel$}. Furthermore, though the hierarchies
$\bigcup_{k}\mbox{P-mc}(k)$ and $\bigcup_{k}\mbox{S}(k)$ are properly
infinite, they both are still contained in
P/poly~\cite{ogi:j:comparable,hem-jia-rot-wat:t:multiselectivity}.
Among a number of other results, all the relations between the classes
$\mbox{P-mc}(j)$ and $\mbox{S}(k)$ are completely established
in Hemaspaandra
et al.~\cite{hem-jia-rot-wat:t:multiselectivity}. These relations are
stated in Lemma~\ref{lem:relations} below, as they'll be referred to
in the upcoming proof of Theorem~\ref{thm:s2-el2}.

\begin{lemma}
\label{lem:relations}
\cite{hem-jia-rot-wat:t:multiselectivity}
\quad
\begin{enumerate}
\item $\mbox{P-mc}(2) \not\seq \bigcup_{k \geq 1}\mbox{S}(k)$.

\item For each $k\geq 1$, $\mbox{S}(k) \subset \mbox{P-mc}(k+1)$\ and \
$\mbox{S}(k) \not\seq \mbox{P-mc}(k)$.\footnote{%
\protect\singlespacing%
This generalizes to $k$ larger than 1 a result of Ogihara who proves
that the P-selective sets are strictly contained in
$\mbox{P-mc}(2)$~\cite{ogi:j:comparable} as well as the known fact
that P-Sel is strictly larger than~P~\cite{sel:j:pselective-tally}.
}%
\end{enumerate}
\end{lemma}

The following result establishes a structural difference between
Selman's P-selectivity and the generalized selectivity introduced
above: Though clearly \mbox{$\mbox{S}(1) = \psel \seq
  \mbox{EL}_{2}$}~\cite{ami-bei-gas:j-subm:uni} and \mbox{$\mbox{\rm
    P-mc}(1) = \p \seq \mbox{EL}_{2}$}, we show that there are sets
(indeed, sparse sets) in $\mbox{S}(2) \cap \mbox{P-mc}(2)$ that are
not in $\mbox{EL}_{2}$.  Previously, Allender and
Hemaspaandra~\cite{all-hem:j:low} have shown that \ppoly\ (and indeed
SPARSE and coSPARSE) is not contained in $\mbox{EL}_{2}$.
Theorem~\ref{thm:s2-el2} and Corollary~\ref{cor:s2-el2}, however,
extend this result and give the first known (and optimal)
$\mbox{EL}_{2}$ lower bounds for generalized selectivity-like classes.

\begin{thm}
\label{thm:s2-el2} \quad
$\mbox{\rm SPARSE} \cap \mbox{\rm S}(2) \cap \mbox{\rm P-mc}(2)
\not\seq \mbox{\rm EL}_{2}$.
\end{thm}

\noindent
{\bf Proof.}  
\quad
Let $t$ be the function defined in the proof of 
Theorem~\ref{thm:el2-join} that gives triple-exponentially spaced gaps. 
Let $T_k  \equalsdef 
\Sigma^{t(k)}$, for $k \geq 0$, and $T  \equalsdef  \bigcup_{k \geq 0} T_k$.
Let EE\label{ind:ee} 
be defined as $\bigcup_{c\ge 0} \mbox{DTIME}[2^{c2^{n}}]$.  We
will construct a set $B$ such that
{\singlespacing
\begin{description}
\item[(a)] $B \seq T$, 

\item[(b)] $B\in \mbox{EE}$, 

\item[(c)] $\| B \cap T_k \| \leq 1$ for each $k \geq 0$, and

\item[(d)] $B \not\in \mbox{EL}_{2}$.
\end{description}
} %

Note that it follows from (a), (b), and (c) that $B$ is a sparse set
in~$\mbox{S}(2)$.  Indeed, any input to the $\mbox{S}(2)$-selector for
$B$ that is not in $T$ (which can easily be checked) is not in $B$
by~(a) and may thus be ignored.  If all inputs that are in $T$ are in
the same $T_k$ then, by~(c), the $\mbox{S}(2)$-promise (that $B$
contains at least two of the inputs) is never satisfied, and the
selector may thus output an arbitrary input.  On the other hand, if
the inputs that are in $T$ fall in more than one~$T_k$, then for all
inputs of length smaller than the maximum length, it can be decided by
brute force whether or not they belong to~$B$---this is possible, as
$B \in \mbox{EE}$ and the $T_k$ are triple-exponentially spaced.  From
these comments, the action of the $\mbox{S}(2)$-selector is clear.

\smallskip

By Lemma~\ref{lem:relations}, $B$ is thus in $\mbox{P-mc}(k)$ for each
$k \geq 3$. But since $\mbox{S}(2)$ and $\mbox{P-mc}(2)$ are
incomparable (again by Lemma~\ref{lem:relations}), we still must argue
that \mbox{$B \in \mbox{P-mc}(2)$}. Again, this follows from (a), (b),
and~(c), since for any fixed two inputs, $u$ and~$v$, if they are of
different lengths, then the smaller one can be solved by brute force;
and if they have the same length, then it is impossible by~(c) that
\mbox{$(\chi_B(u), \chi_B(v)) = (1,1)$}. In each case, one out of the
four possibilities for the membership of~$u$ and~$v$ in~$B$ can be
excluded in polynomial time. Hence, $B \in \mbox{P-mc}(2)$.

\smallskip

For proving (d), we will construct $B$ such that $\np^B \not\seq
\conp^{B \oplus \mbox{\protect\scriptsize SAT}}$ (which clearly implies that
$\np^{\mbox{\protect\scriptsize NP}^{B}} \not\seq 
\np^{B \oplus \mbox{\protect\scriptsize
    SAT}}$).  Define
\[
L_B  \equalsdef  \{ 0^n \,|\, (\exists x)\, [ |x| = n\ \wedge\  x \in B] \}.
\]
Clearly, $L_B \in \np^B$. As in the proof of 
Theorem~\ref{thm:el2-join}, let $\{ N_i \}_{i\geq 1}$ be a standard
enumeration of all \conp\ oracle machines satisfying the condition
that the runtime of each $N_i$ is independent of the oracle and each
machine is repeated infinitely often in the enumeration.
Let $p_i$ be the polynomial bound on the runtime of $N_i$.
The set $B  \equalsdef  \bigcup_{i\geq 0} B_i$ is constructed in stages.  In
stage $i$, at most one string of length $n_i$ will be added to $B$,
and $B_{i-1}$ will have previously been set to the content of $B$ up
to stage~$i$.  Initially, $B_0 = \emptyset$ and $n_0 = 0$. Stage $i >
0$ is as follows: 

\smallskip

Let $n_i$ be the smallest number such that (i)~$n_i > n_{i-1}$,
(ii)~$n_i = t(k)$ for some~$k$, and (iii)~$2^{n_i} > p_i(n_i)$.
Simulate $N_{i}^{B_{i-1} \oplus \mbox{\protect\scriptsize
    SAT}}(0^{n_i})$.
\begin{description}
\item[Case 1:] If $N_{i}^{B_{i-1} \oplus \mbox{\protect\scriptsize
      SAT}}(0^{n_i})$ rejects (in the sense of~coNP, i.e., if there
  are one or more rejecting computation paths), then fix some
  rejecting path and let $w_i$ be the smallest string of length $n_i$
  that is not queried along this path. Note that, by our choice of
  $n_i$, such a string $w_i$, if needed, must always exist. Set
  \mbox{$B_i := B_{i-1} \cup \{ w_i \}$}.

\item[Case 2:] If $0^{n_i} \in L(N_{i}^{B_{i-1} \oplus
  \mbox{\protect\scriptsize SAT}})$, then set \mbox{$B_i := B_{i-1}$}.

\item[Case 3:] If the simulation of $N_{i}^{B_{i-1} \oplus
    \mbox{\protect\scriptsize SAT}}$ on input $0^{n_i}$ fails to be
  completed in double exponential (say, $2^{100 \cdot 2^{n_i}}$ steps)
  time (for example, because $N_i$ is huge in size relative to $n_i$),
  then abort the simulation and set \mbox{$B_i := B_{i-1}$}.
\end{description}
This completes the construction of stage~$i$.  

\smallskip

Since we have chosen an enumeration such that the same machine as
$N_i$ appears infinitely often and as the $n_i$ are strictly
increasing, it is clear that for only a finite number of the $N_{i_1},
N_{i_2}, \ldots $ that are the same machine as $N_i$ can Case~3 occur
(and thus $N_i$, either directly or via one of its clones, is
diagonalized against eventually).  Note that the construction meets
requirements (a), (b), and (c) and shows $L_B \neq L(N_{i}^{B \oplus
  \mbox{\protect\scriptsize SAT}})$ for each $i\geq 1$.~\hfill$\Box$

\medskip

Since $\mbox{EL}_2$ and $\mbox{\rm P-mc}(2)$ are both closed under
complementation, we have the following corollary.

\begin{cor}
\label{cor:s2-el2}
\quad
$\mbox{\rm coSPARSE} \cap \mbox{\rm coS}(2) \cap
\mbox{P-mc}(2) \not\seq \mbox{\rm EL}_{2}$.
\end{cor}

When suitably combined with standard techniques of constructing
P-selective sets, the proof of the previous theorem even proves that
the second level of the extended low hierarchy is not closed under a
number of Boolean operations, as we have claimed in the beginning of
this section. These results extend the main result
of Hemaspaandra and 
Jiang~\cite{hem-jia:j:psel} 
which says that P-Sel is not
closed under those Boolean connectives. 

Let us first adopt and slightly generalize some of the formalism used
in~\cite{hem-jia:j:psel} so as to suffice for our
objective.  The intuition is that we want to show that certain {\em
  widely-spaced\/} and {\em complexity-bounded\/} sets whose 
definition will be based on the set
$B$ constructed in the previous proof are P-selective. Fix some
complexity-bounding function $f$ and some wide-spacing function $\mu$
such that the spacing is at least as wide as given by the following
inductive definition: $\mu(0)=2$ and \mbox{$\mu(i+1) = 2^{f(\mu(i))}$}
for each $i\ge 0$. Now define for each~$k \geq 0$,
\[
R_k  \equalsdef  \{ i \,|\, i \in \N\, \wedge\,  \mu(k)\le i < \mu(k+1)\}, 
\]
and the following two classes of languages (where we will implicitly
use the standard correspondence between $\sigmastar$ and $\N$): 
\[{\cal C_1}  \equalsdef  \{ A \seq \N \,|\, (\forall j\ge 0)\, 
[R_{2j}\cap A = 
\emptyset\,\wedge\, (\forall x,y \in R_{2j+1})\, [(x\le y \,\wedge\, x\in A )
\, \Longrightarrow  \,  y\in A]]\};
\]
\[
{\cal C_2} \equalsdef \{ A \seq \N \,|\, (\forall j\ge 0)\,
[R_{2j}\cap A = \emptyset\,\wedge\, (\forall x,y \in R_{2j+1})\,
[(x\le y \,\wedge\, y\in A) \, \Longrightarrow \, x\in A]]\}.
\]
 
In~\cite{hem-jia:j:psel}, the following lemma is
proven for the particular complexity-bounding function $f'(n) =
2^{{\cal O}(n)}$ and for the classes ${\cal C}^{'}_{1}$ and ${\cal
  C}^{'}_{2}$ having implicit in their definition the particular
wide-spacing function that is given by $\mu'(0)=2$ and
\mbox{$\mu'(i+1) = 2^{2^{(\mu'(i))}}$}, $i \geq 0$.  However, there is
nothing special about these functions $f'$ and~$\mu'$, i.e., for
Lemma~\ref{lem:hemjia} to hold it suffices that $f$ and $\mu$ relate
to each other as required above.  In light of this, the proof of
Lemma~\ref{lem:hemjia} is quite analogous to the proof given
in~\cite{hem-jia:j:psel}.

\begin{lemma}
\label{lem:hemjia}
\quad
${\cal C_1} \cap \mbox{DTIME}[f] \seq \psel\ $ and
$\ {\cal C_2} \cap \mbox{DTIME}[f] \seq \psel$.
\end{lemma}

Now we are ready to prove the main result of this section. 

\begin{thm}
\label{thm:el2-not-closed}
\quad
$\mbox{\rm EL}_{2}$ is not closed under intersection, union, exclusive-or, or 
equivalence.
\end{thm}

\noindent
{\bf Proof.} \quad
Using the technique of~\cite{hem-jia:j:psel}, 
it is not hard to prove that the set $B$
constructed in the proof of Theorem~\ref{thm:s2-el2} can be represented 
as $B = A_1 \cap A_2$ for P-selective sets $A_1$ and $A_2$. 
More precisely, let
\begin{eqnarray*}
A_1 &  \equalsdef  & \{x \,|\, (\exists w \in B)\, [ |x| = |w| \,\wedge\, 
x \leq_{\mbox{\protect\scriptsize lex}} w ] \},\\
A_2 &  \equalsdef  & \{x \,|\, (\exists w \in B)\, [ |x| = |w| \,\wedge\, 
w \leq_{\mbox{\protect\scriptsize lex}} x] \}.
\end{eqnarray*}
Since $B\in \mbox{EE}$ and is triple-exponentially spaced, we have
from Lemma~\ref{lem:hemjia} that $A_1$ and $A_2$ are in P-Sel and thus
in~$\mbox{EL}_2$. On the other hand, we have seen in the previous proof
that $B = A_1 \cap A_2$ is not in $\mbox{EL}_2$. Similarly, if we
define
\begin{eqnarray*}
C_1 &  \equalsdef  & \{x \,|\, (\exists w \in B)\, [ |x| = |w| \,\wedge\, 
x <_{\mbox{\protect\scriptsize lex}}  w ] \},\\
C_2 &  \equalsdef  & \{x \,|\, (\exists w \in B)\, [ |x| = |w| \,\wedge\, 
x \leq_{\mbox{\protect\scriptsize lex}} w] \},
\end{eqnarray*}
we have $B = C_1 \ \xor\ C_2$, where $\xor$ denotes the exclusive-or
operation.  As before, $C_1$ and $C_2$ are in P-Sel and thus
in~$\mbox{EL}_2$.  Hence, $\mbox{EL}_2$ is not closed under
intersection or exclusive-or.  Since $\mbox{EL}_2$ is closed under
complementation, it must also fail to be closed under union and
equivalence.~\hfill$\Box$

\medskip

The proof of the above result also gives the following corollary.

\begin{corollary}
\label{cor:el2-not-closed}
{\rm \cite{hem-jia:j:psel}}
\quad
\psel\ is not closed under intersection, union, exclusive-or, or 
equivalence.
\end{corollary}

{\samepage
\begin{center}
{\bf Acknowledgments}
\end{center}
\nopagebreak
\noindent
We thank an anonymous referee for stressing that our
results should be interpreted as evidence regarding the 
unnaturalness of
extended lowness as a complexity measure.

}%

\bibliographystyle{alpha}

{\singlespacing \bibliography{main.bbl}}

\end{document}